\documentstyle[12pt]{article}
\newlength{\extraspace}
\newlength{\extraspaces}
\newcommand{\be}{\begin{equation}
\addtolength{\abovedisplayskip}{\extraspaces}
\addtolength{\belowdisplayskip}{\extraspaces}
\addtolength{\abovedisplayshortskip}{\extraspace}
\addtolength{\belowdisplayshortskip}{\extraspace}}
\newcommand{\ee}{\end{equation}}

\newcommand{\ba}{\begin{eqnarray}
\addtolength{\abovedisplayskip}{\extraspaces}
\addtolength{\belowdisplayskip}{\extraspaces}
\addtolength{\abovedisplayshortskip}{\extraspace}
\addtolength{\belowdisplayshortskip}{\extraspace}}
\newcommand{\ea}{\end{eqnarray}}

\newcommand{\newsection}[1]{
\vspace{12mm}
\pagebreak[3]
\addtocounter{section}{1}
\setcounter{equation}{0}
\setcounter{subsection}{0}
\setcounter{footnote}{0}
\noindent{\bf \thesection. #1}
\nopagebreak
\medskip
\nopagebreak}

\newcommand{\newsubsection}[1]{
\vspace{0.8cm}
\pagebreak[3]
\addtocounter{subsection}{1}
\noindent{\it \thesubsection. #1}
\nopagebreak
\vspace{2mm}
\nopagebreak}

\begin{document}
\addtolength{\baselineskip}{1.5mm}

\thispagestyle{empty}
\begin{flushright}
hep-th/9706101
\end{flushright}
\vbox{}
\vspace{0.5cm}

\begin{center}
{\LARGE{Monopoles, vortices and kinks \\[2mm]
        in the framework of non-commutative geometry}}\\[16mm]
{Edward Teo}\\[6mm]
{\it Department of Applied Mathematics and Theoretical Physics,
University of Cambridge,\\[1mm]
Silver Street,
Cambridge CB3 9EW,
England}\\[3mm]

and \\[3mm]

{\it Department of Physics, 
National University of Singapore,
Singapore 119260}\\[15mm]

{Christopher Ting}\\[6mm]
{\it Department of Computational Science,
National University of Singapore,
Singapore 119260}
\end{center}
\vspace{1.3cm}

\centerline{\bf Abstract}\bigskip
\noindent
Non-commutative differential geometry allows a scalar field to be regarded 
as a gauge connection, albeit on a discrete space.  We explain how the 
underlying gauge principle corresponds to the independence of physics on 
the choice of vacuum state, should it be non-unique.  A consequence is that 
Yang--Mills--Higgs theory can be reformulated as a generalised Yang--Mills 
gauge theory on Euclidean space with a ${\bf Z}_2$ internal structure.  By 
extending the Hodge star operation to this non-commutative space, we are 
able to define the notion of self-duality of the gauge curvature form in 
arbitrary dimensions.  It turns out that BPS monopoles, critically coupled 
vortices, and kinks are all self-dual solutions in their respective 
dimensions.  We then prove, within this unified formalism, that static 
soliton solutions to the Yang--Mills--Higgs system exist only in one, two 
and three spatial dimensions.



\newpage

\newsection{Introduction}

A theory of scalar fields, possessing some symmetry, is said to undergo 
{\it spontaneous symmetry breaking\/} when its potential acquires a family 
of degenerate minima.  In this event, the vacuum of the theory is not 
unique, and possible states are related through the symmetry.  Any specific 
choice of a vacuum state however, breaks this invariance.

Scalar fields with spontaneously broken symmetry play an important r\^ole 
in modern Yang--Mills gauge theory.  Through the Higgs mechanism 
\cite{Higgs}, they generate masses for the gauge bosons as required by 
phenomenology.  Baring its experimental success however, the form and 
content of the Higgs sector lacks motivation from gauge principles so vital 
to the corresponding Yang--Mills sector.  Because of this, it is often 
regarded as an {\it ad hoc\/} and aesthetically unappealing feature in the 
otherwise geometrically beautiful backdrop of Yang--Mills theory, and there 
has been several attempts in the literature to address this problem.

In the early 1980s, several authors \cite{Fairlie,Manton1} proposed a 
Kaluza--Klein unification of Yang--Mills and Higgs fields.  Gauge theory 
was formulated on a higher-dimensional space-time, and components of the 
gauge connection in the extra dimensions identified as Higgs fields.  A 
dimensional reduction yielded four-dimensional Yang--Mills theory together 
with a symmetry-breaking potential for the Higgs fields.  These models 
offered predictions for otherwise free parameters like the Higgs mass, but 
failed to reproduce the standard model of electroweak interactions.

It was more recently realised that one could replace the higher-dimensional 
spaces of these Kaluza--Klein theories by discrete structures 
\cite{Madore1}, on which a generalised notion of differential geometry can 
be set up using non-commutative geometry 
\cite{Connes1,Connes3,Varilly,Cham1,Cham2}.  
Already, the simplest case of space-time with an internal structure 
consisting of two points is sufficient to exhibit the desired behaviour 
\cite{Connes2,Coquereaux2}.  Gauge theory formulated on this extended geometry 
yields a connection form, whose internal component between the two points 
can be interpreted as a scalar field on space-time.  We may then define an 
extended curvature form, and construct gauge-invariant actions from it.  
The usual choice leads to Yang--Mills theory and a Higgs field transforming 
in the adjoint representation, with the familiar symmetry-breaking 
potential appearing naturally.  In contrast to the Kaluza--Klein models, an 
arbitrary truncation of continuous space degrees of freedom is now 
unnecessary.

More complicated Higgs sectors have been derived by generalising the 
two-point space to other discrete structures \cite{Madore2}.  There has 
also been much effort devoted to constructing a realistic model 
\cite{Balakrishna,Coquereaux3,Martin}, that would reproduce the 
standard electroweak theory.  It was hoped such a formulation of the 
Yang--Mills--Higgs system was constrained enough to predict a classical 
value for the Higgs mass \cite{Connes1}, thus raising the possibility of 
experimental verification.  However, it is now generally accepted that 
there is enough freedom in the theory to make `different' numerical 
predictions.  As emphasised in ref.\ \cite{Coquereaux3}, what emerges is 
precisely the standard electroweak model, with not one free parameter less.  
It seems non-commutative differential geometry just furnishes a new and 
somewhat arcane way to rewrite models of particle physics.

Yang--Mills--Higgs theories are also important in that they admit a variety 
of soliton solutions whose existence and stability are due to topological 
factors.  They are non-dissipative, finite-energy field configurations 
possessing boundary conditions at infinity which are topologically 
different from those of a vacuum.  The prototype is the 't~Hooft--Polyakov 
monopole in three spatial dimensions \cite{tHooft,Polyakov}.  Its 
two-dimensional analogue are vortices, discovered by Nielsen and Olesen 
\cite{Nielsen}.  In one dimension where the Yang--Mills sector becomes 
trivial, an example is the kink solution of $\varphi^4$ 
theory.\footnote{Strictly speaking, the kink is not a soliton but a 
`solitary wave' (see for example, ref.~\cite{Rajaraman}).  This difference 
need not concern us here.}

Each of these solutions has an energy which is bounded from below by a 
non-zero quantity depending only on the topology or boundary conditions of 
the system, hence its stability against decay to the vacuum.  This 
so-called Bogomol'nyi bound is saturated by field configurations satisfying 
certain first-order equations, which imply the full equations of motion but 
are in general much easier to solve and analyse \cite{Bogomolnyi}.  
Monopoles in the BPS limit \cite{Prasad} and vortices with critical 
coupling are examples when the energy is minimised.

The most well-known example of this in fact comes from pure Yang--Mills 
theory in a four-dimensional Euclidean space.  Denote by $F$ the gauge 
curvature two-form and $\ast F$ its Hodge dual.  The non-negativity of the 
inner product $(F\mp\ast F,F\mp\ast F)$ implies the inequality
\be
(F,F)\geq|(F,\ast F)|\,,
\ee
the left-hand side being the action functional or energy of the system.  
Equality occurs if and only if the curvature is self-dual or 
anti-self-dual:
\be
\label{duality}
\ast F=\pm F\,.
\ee
Solutions to this set of first-order equations are known as Yang--Mills 
instantons \cite{Belavin}, and their energy is proportional to the 
Pontryagin index of the geometry.

Unfortunately, this elegant geometrical interpretation has found limited 
application to Yang--Mills--Higgs systems.  In the case of BPS monopoles, 
one can regard the Higgs field $\varphi$ as the last component of a {\it 
four\/}-dimensional gauge connection $(A,\varphi)$.  The self-duality 
condition (\ref{duality}) is then equivalent to the Bogomol'nyi equations 
governing these monopoles \cite{Manton2}.  Such an argument has not been 
extended to other examples like vortices, as this construction is peculiar 
to four-dimensional systems.

The aim of the present paper is to generalise this type of self-duality 
property to Yang--Mills--Higgs systems in arbitrary dimensions.  To do so, 
we exploit the fact that scalar fields with a $\varphi^4$ potential can be 
reformulated as a two-dimensional Yang--Mills theory on a discrete ${\bf 
Z}_2$ geometry.  A Yang--Mills--Higgs theory in $n$ Euclidean dimensions is 
therefore equivalent to a pure $(n+2)$-dimensional Yang--Mills theory on 
the Euclidean space with a ${\bf Z}_2$ internal structure.  We define a 
generalised notion of Poincar\'e duality on this non-commutative geometry, 
as well as what is meant by self-duality of the gauge curvature form in 
arbitrary spatial dimensions.  It is then shown that critically coupled 
vortices and kinks are self-dual solutions in the extended sense.  BPS 
monopoles and instantons also emerge as self-dual examples of this unified 
formalism.

We should point out there is a rather loose sense of `self-duality' in use 
in the literature (see for example, ref.\ \cite{Dunne}).  It refers to 
theories which have special interactions and coupling strengths, such that 
the second-order equations of motion reduce to first-order ones.  Solutions 
in general minimise a Bogomol'nyi-type bound for the energy.  It is in this 
sense that critically coupled vortices have hitherto been referred to as 
self-dual.  Our notion of self-duality is a more exacting one which 
requires the concept of a Hodge star operation, as in (\ref{duality}), on 
the appropriate geometry.

We begin by reviewing how differential geometry on the ${\bf Z}_2$ geometry 
can be set up, as per Coquereaux~{\it et.~al.\/} \cite{Coquereaux2}.  In 
doing so, we shall recast the formalism, as far as possible, in a language 
that makes it (formally) similar to ordinary differential geometry.  This 
is in line with the general philosophy of Madore~{\it et.~al.\/} 
\cite{Madore1, Madore}, and would lead to two new results.  The first is we 
are able to define a Hodge star operator for this case, in direct analogy 
with the usual one.  The second is we can now identify the Maurer--Cartan 
one-form $\theta$ of this geometry.  Unlike the ordinary case, it has a 
non-vanishing curvature which can ultimately be identified as the Higgs 
mass \cite{Madore3}.  Using this interpretation, we explain how the gauge 
principle underlying Higgs fields corresponds to the independence of 
physics on the choice of vacuum state.

In the second part of the paper, we consider some explicit examples to 
which this formalism can be applied.  In particular, kinks in $n=1$ spatial 
dimension, vortices in $n=2$, monopoles in $n=3$ and instantons in $n=4$ 
are studied.  We show that the appropriate self-duality condition on the 
generalised curvature form $\Omega$ takes the general form
\be
\ast\Omega\simeq\pm\Omega\wedge\theta^{n-2},
\ee
and that its solutions minimise the energy functional.  The corresponding 
topological bound is calculated for each case, and verified to be in 
agreement with standard results.  We finally prove, within our unified 
formalism, that static soliton solutions to the Yang--Mills--Higgs system 
can only exist when $n\leq3$ \cite{Goddard, Jaffe}.  This is a 
generalisation of the well-known result for pure Yang--Mills theory 
\cite{Deser}, which states that $n=4$ is the only dimension in which 
solitons are allowed.

\newsection{Differential geometry on ${\bf Z}_2$}

\newsubsection{Differential forms}

Consider the cyclic group of order two, ${\bf Z}_2=\{e,r\,|\,r^2=e\}$, 
which has the explicit matrix representation \cite{Landi}
\be
\pi(e)=\pmatrix{1&0\cr0&1},\qquad \pi(r)=\pmatrix{1&0\cr0&-1}.
\ee
The algebra of complex functions on ${\bf Z}_2$ can therefore be realised 
as the algebra ${\cal M}_2^+$ of {\it diagonal\/} $2\times2$ matrices, with 
the usual rules of addition and multiplication.  It is a subalgebra of the 
algebra ${\cal M}_2$ of complex $2\times2$ matrices generated by the Pauli 
matrices $\tau_i$.

The exterior derivative of an element $a\in {\cal M}_2^+$ is defined 
to be the commutator
\be
{\rm d}a=im\,[\eta,a]\,,
\ee
where $m$ is a mass scale and $\eta$ a real linear combination of $\tau_1$ 
and $\tau_2$.  It is actually sufficient to set
\be
\label{eta}
\eta=\cos\gamma\,\tau_1+\sin\gamma\,\tau_2\,,
\ee
for some angle parameter $\gamma$, since a global factor can be absorbed 
into $m$~\cite{Coquereaux2}.  Note that the usual 
Leibniz rule is satisfied by this definition.

A one-form on ${\bf Z}_2$ has the general form $a\,{\rm d}b$, where 
$a,b\in{\cal M}_2^+$, and is an off-diagonal $2\times2$ matrix in the 
present representation.  A basis for the space of one-forms is supplied by
\be
\label{one-form}
\theta^1=\hbox{$\frac{1}{2}$}{\rm d}\tau_3\,,
\ee
although another possible choice is
\be
\label{one-form2}
\theta^2=i\tau_3\theta^1.
\ee
Together, they span the space over the complex numbers ${\bf C}$.

The exterior derivative of a one-form $\alpha$ on ${\bf Z}_2$ is given 
by the anti-commutator
\be
\label{anti-comm}
{\rm d}\alpha=-m\,\{\eta,\alpha\}\,.
\ee
The nilpotency condition ${\rm d}^2=0$ follows from the identity 
$\eta^2={\bf 1}$.  Further requiring ${\rm d}\tau_3\wedge{\rm d}\tau_3 
={\rm d}(\tau_3\,{\rm d}\tau_3)$ to hold implies that the wedge product 
between one-forms $\alpha$ and $\beta$ is their matrix product, with an 
extra factor of $i$:
\be
\label{wedge1}
\alpha\wedge\beta=i\alpha\beta\,.
\ee
The graded Leibniz rule, given by ${\rm d}(a\alpha)={\rm d}a\wedge\alpha
+a{\rm d}\alpha$ and ${\rm d}(\alpha a)={\rm d}\alpha a
-\alpha\wedge{\rm d}a$, applies.

The space of two-forms on ${\bf Z}_2$ consists of elements $a\,{\rm d}b 
\wedge{\rm d}c$, where $a,b,c\in{\cal M}_2^+$.  Any two-form $\Omega$ can be 
written as
\be
\label{two-form}
\Omega=\Omega_{12}\,\theta^1\wedge\theta^2,
\ee
with $\Omega_{12}\in{\cal M}_2^+$.  Note that $\theta^1\wedge\theta^2 
=-\,\theta^2\wedge\theta^1=m^2\tau_3$.  This means the space of two-forms 
is isomorphic to the algebra ${\cal M}_2^+$ itself.  The universal algebra 
of forms on ${\bf Z}_2$ is therefore ${\cal M}_2$.  It has a graded 
structure, whereby forms of even degree are diagonal matrices and forms of 
odd degree are off-diagonal.  The exterior derivative operator takes even 
forms into odd ones, and {\it vice versa\/}.  We shall extend the 
definition of the wedge product (\ref{wedge1}) so that it denotes matrix 
multiplication between an even and an odd form, or between two even forms.

The above definitions of the exterior derivative operator have been chosen 
so it satisfies
\be
({\rm d}a)^\dagger={\rm d}a^\dagger,
\ee
for any $a\in{\cal M}_2$, where the involution $\dagger$ denotes Hermitian 
conjugation.  As a consequence, $\theta^1$ and $\theta^2$ are Hermitian.  
We also have
\be
(a\wedge b)^\dagger=(-1)^{{\rm deg}\,a\,{\rm deg}\,b}\,
b^\dagger\wedge a^\dagger,
\ee
where $a,b\in{\cal M}_2$. These Hermiticity
properties are identical to those obeyed by ordinary differential forms.

The one-form $\theta=-\theta^2$ will turn out to play an important r\^ole 
in the construction of a gauge theory below.  The exterior derivative of 
$a\in{\cal M}_2$ can be rewritten in terms of $\theta$ as~\cite{Madore1}
\be
\label{exterior}
{\rm d}a=-i[\theta,a]\,,
\ee
where $[a,b]=a\wedge b-(-1)^{{\rm deg}\,a\,{\rm deg}\,b}\,b\wedge a$ 
henceforth denotes the {\it graded\/} commutator.  It respects the graded 
Leibniz rule
\be
{\rm d}(a\wedge b)={\rm d}a\wedge b+(-1)^{{\rm deg}\,a}a\wedge{\rm d}b\,.
\ee
Observe that (\ref{exterior}) implies the identity
\be
\label{Maurer}
{\rm d}\theta+i\,\theta\wedge\theta=m^2.
\ee
$\theta$ is the analogue of the Maurer--Cartan form in ordinary 
differential geometry (see for example, ref.~\cite{Eguchi}).  We shall 
return to this fact in sec.~3.3.

\newsubsection{Metric structure}

There is a natural metric structure on ${\bf Z}_2$ given by 
\cite{Madore1,Madore}
\be
g_{ab}=\hbox{$\frac{1}{2m^4}$}{\rm Tr}\,(\theta^a{}^\dagger\theta^b)\,,
\ee
where $a,b=1,2$.  $g_{ab}$ has been normalised so that it is the 
two-dimensional Euclidean metric, with physical dimensions of inverse-mass 
squared.

A Hodge $\ast$ operation can be defined on the universal algebra of forms 
on ${\bf Z}_2$ by a straightforward application of the usual formula 
\cite{Eguchi}
\be
\label{star}
\ast(\theta^{a_1}\wedge\cdots\wedge\theta^{a_p})={1\over(n-p)!}
\sqrt{g}\,\epsilon^{a_1\cdots a_p}{}_{a_{p+1}\cdots a_n}\,\theta^{a_{p+1}}
\wedge\cdots\wedge\theta^{a_n}.
\ee
Here, $n=2$ is the dimension of the geometry in question, $\epsilon_{a_1 
\cdots a_n}$ the $n$-dimensional anti-symmetric tensor, and 
$g=\frac{1}{m^4}$ the determinant of the metric.  Indices are raised and 
lowered with $g_{ab}$.  Note that $\ast$ maps $p$-forms into $(n-p)$-forms, 
hereby extending the notion of Poincar\'e duality to this case.  The 
even--odd grading of the universal algebra is preserved under this duality 
transformation.  Explicitly, we have $\ast\,{\bf 
1}=\sqrt{g}\,\theta^1\wedge\theta^2$, $\ast\,\theta^1=\theta^2$, and their 
inverse relations which follow from the property $\ast^2=(-1)^{p(n-p)}$.

An invariant volume element on ${\bf Z}_2$ is provided by 
$\sqrt{g}\,\theta^1\wedge\theta^2$.  We define the integral of a two-form 
$\Omega$ to be its matrix supertrace 
\cite{Coquereaux2}:
\be
\int\Omega={\rm STr}\,\Omega\,.
\ee
The appearance of the supertrace should not be surprising as we are dealing 
with graded matrices.  It is equivalent to the formula
\be
\int a\sqrt{g}\,\theta^1\wedge\theta^2={\rm Tr}\,a\,.
\ee
An inner product on ${\cal M}_2$, which immediately follows, is
\be
(a,b)=\int a^\dagger\wedge\ast b\,.
\ee
It is identical to the one usually adopted for complex matrices.

\newsubsection{Gauge theory}

Let ${\cal U}=\{g\in{\cal M}_2^+\,|\,g^\dagger g=1\}$ be the group of 
unitary elements of ${\cal M}_2^+$.  We would like to construct a gauge 
theory on ${\bf Z}_2$, with ${\cal U}$ as the group of symmetry 
transformations.  Multiplying a function on ${\bf Z}_2$ by $g\in{\cal 
U}$ corresponds to performing two global ${\rm U}(1)$ transformations, 
one for each element of ${\bf Z}_2$.

This gauge symmetry is a local one, since, in general, ${\rm d}g\neq0$.  As 
in ordinary gauge theory, we have to introduce a covariant derivative ${\rm 
D}={\rm d}+i[\omega,\,\cdot\,]$, where $\omega$ is a Hermitian one-form 
known as the gauge connection.  We require it to gauge-transform 
covariantly under the adjoint action of ${\cal U}$, namely ${\rm 
D}\rightarrow g^{-1}{\rm D}g$, so that
\be
\label{g_transf}
\omega~\rightarrow~g^{-1}\omega g-g^{-1}i{\rm d}g\,.
\ee
The curvature two-form is then defined to be
\be
\label{curvature}
\Omega={\rm d}\omega+i\,\omega\wedge\omega\,.
\ee
It is Hermitian and transforms covariantly under (\ref{g_transf}).

We shall write the gauge connection as \cite{Madore1}
\be
\label{connection-form}
\omega=\theta+\phi\,,
\ee
with $\theta$ as the preferred origin.  The gauge transformation of $\theta$ 
is defined to be
\be
\theta~\rightarrow~g^{-1}\theta g-g^{-1}i{\rm d}g\,.
\ee
But since the right-hand side equals the left by (\ref{exterior}), $\theta$ 
is in fact a {\it gauge-invariant\/} quantity.  $\phi$ therefore transforms 
covariantly as
\be
\phi~\rightarrow~g^{-1}\phi g\,.
\ee
In this case, the covariant derivative of an element $a\in{\cal M}_2$ 
takes the form
\be
{\rm D}a={\rm d}a+i[\omega,a]=i[\phi,a]\,.
\ee
Using the identity (\ref{Maurer}), we see the curvature form 
(\ref{curvature}) is explicitly
\be
\label{curvature-form1}
\Omega=m^2-\phi^2.
\ee
Thus, the term on the right-hand side of (\ref{Maurer}) is non-zero because 
$\theta$ has curvature.  Because it is gauge invariant, we cannot make 
$\theta$ vanish by a choice of gauge.  Note that the Bianchi identity ${\rm 
D}\Omega=0$ is trivially satisfied.

The usual starting point for the study of Yang--Mills theory is the action, 
normally taken to be the norm-square of the curvature form.  In the present 
case, such a term is
\be
(\Omega,\Omega)={\rm Tr}\,(\Omega_{12}^\dagger\Omega^{12})\,.
\ee
It is clearly invariant under the gauge transformation (\ref{g_transf}).  
There is however, another possible choice, absent in the Yang--Mills case, 
given by\footnote{In the Yang--Mills case, the analogue of this term is the 
contraction of the metric tensor $g_{\alpha\beta}$ with the Yang--Mills 
field strength $F_{\alpha\beta}$, which, of course, vanishes identically.}
\be
|(\theta\wedge\theta,\Omega)|={\rm Tr}\,(\epsilon^{12}\Omega_{12}\tau_3)\,.
\ee
That such a term should not be ignored in non-commutative geometry was 
pointed out by Sitarz \cite{Sitarz1,Sitarz2}.

\newsection{Yang--Mills--Higgs theory}

\newsubsection{Differential geometry on $M\times{\bf Z}_2$}

The geometry of interest in this paper is an $n$-dimensional Euclidean 
space $M$ with a ${\bf Z}_2$ internal structure.  The algebra of functions 
on this extended geometry is the tensor product of the algebra ${\cal 
M}_2^+$, introduced in the preceding section, with the algebra ${\cal C}$ 
of complex functions on $M$.  An element of ${\cal M}_2^+\otimes{\cal C}$ 
has the explicit form
\be
\pmatrix{f_1&0\cr0&f_2}, 
\ee
where $f_1$ and $f_2$ are functions on $M$.

Let $a$ be a form on ${\bf Z}_2$, and $A$ one on $M$.  We denote by 
$a\otimes A$ a generalised form on $M\times{\bf Z}_2$, with total degree 
${\rm deg}\,a+{\rm deg}\,A$.  The space of generalised one-forms can be 
written as the direct sum~\cite{Madore1}
\be
\Lambda^1=\Lambda^1_{\rm H}\oplus\Lambda^1_{\rm V}\,.
\ee
The so-called horizontal part $\Lambda^1_{\rm H}={\cal 
M}_2^+\otimes\Lambda^1({\cal C})$ consists of diagonal $2\times2$ matrices 
with each component a one-form on $M$, while off-diagonal matrices with 
scalar entries make up the vertical part $\Lambda^1_{\rm V}=\Lambda^1({\cal 
M}_2^+)\otimes{\cal C}$.  Let \{$\theta^\alpha$, $\alpha=1,\dots,n$\} and 
\{$\theta^a$, $a=n+1,n+2$\} be generators of $\Lambda^1_{\rm H}$ and 
$\Lambda^1_{\rm V}$ respectively.  We shall take the former to be the usual 
basis of one-forms ${\rm d}x^\alpha$ on $M$, and latter to be given by 
(\ref{one-form}) and (\ref{one-form2}).  
The complete set of generators of $\Lambda^1$ will be denoted by 
$\theta^i=\{\theta^\alpha,\theta^a\}$.

The wedge product between two forms on this extended geometry is defined 
to be \cite{Coquereaux2}
\be
\label{wedge2}
(a\otimes A)\wedge(a^\prime\otimes A^\prime)
=(-1)^{\deg a^\prime\deg A}\,(a\wedge a^\prime)\otimes(A\wedge A^\prime)\,.
\ee
The wedge product between $A$ and $A^\prime$ is the ordinary wedge product 
between horizontal forms, while the wedge product on ${\bf Z}_2$ is 
understood between $a$ and $a^\prime$.  Forms of higher degree can then be 
systematically constructed from $\theta^i$.  For example, a two-form 
$\Omega$ has the general expansion
\be
\Omega=\hbox{$\frac{1}{2}$}\Omega_{ij}\,\theta^i\wedge\theta^j=
\hbox{$\frac{1}{2}$}\Omega_{\alpha\beta}\,\theta^\alpha\wedge\theta^\beta
+\hbox{$\frac{1}{2}$}\Omega_{\alpha a}\,\theta^\alpha\wedge\theta^a
+\hbox{$\frac{1}{2}$}\Omega_{a\alpha}\,\theta^a\wedge\theta^\alpha
+\hbox{$\frac{1}{2}$}\Omega_{ab}\,\theta^a\wedge\theta^b,
\ee
where $\Omega_{ij}\in{\cal M}_2^+\otimes{\cal C}$. $\Omega_{\alpha\beta}$ 
is the horizontal component of $\Omega$, $\Omega_{ab}$ the vertical 
component, and $\Omega_{\alpha a}$ and $\Omega_{a\alpha}$ the mixed 
components. Note that (\ref{wedge2}) implies the relation 
$\theta^\alpha\wedge\theta^a=-\theta^a\wedge\theta^\alpha$, and so
$\Omega_{\alpha a}=-\Omega_{a\alpha}$.

When written as a matrix, a generalised $n$-form has the form 
\be
\label{matrix_form}
\pmatrix{A+B&C\cr C^\prime&A^\prime+B^\prime},
\ee
where $A$, $A^\prime$ are horizontal $n$-forms, $B$, $B^\prime$ are 
$(n-2)$-forms, while $C$, $C^\prime$ are $(n-1)$-forms.  Thus, the 
components of (\ref{matrix_form}) need not have homogeneous degree, 
although it can be written as a sum of matrices which do.  The wedge 
product (\ref{wedge2}) in terms of such matrices is
\be
\pmatrix{A&C\cr D&B}\wedge
\pmatrix{A^\prime&C^\prime\cr D^\prime&B^\prime}=\pmatrix{
A\wedge A^\prime+(-1)^{{\rm deg}\,C}iC\wedge D^\prime&
C\wedge B^\prime+(-1)^{{\rm deg}\,A}A\wedge C^\prime\cr
D\wedge A^\prime+(-1)^{{\rm deg}\,B}B\wedge D^\prime&
B\wedge B^\prime+(-1)^{{\rm deg}\,D}iD\wedge C^\prime}.
\ee

The exterior derivative operator on $M\times{\bf Z}_2$ can be decomposed 
into a direct sum of its horizontal and vertical parts:
\be
{\rm d}={\rm d}_{\rm H}\oplus{\rm d}_{\rm V}\,,
\ee
where ${\rm d}_{\rm H}=\partial_\alpha\,{\rm d}x^\alpha$ is the ordinary 
exterior derivative operator on $M$, and ${\rm d}_{\rm V}$ is that 
corresponding to ${\bf Z}_2$, given by (\ref{exterior}).  Demanding that 
${\rm d}^2=0$ requires ${\rm d}_{\rm H}$ and ${\rm d}_{\rm V}$ to 
anti-commute.  The exterior derivative of a generalised form is given by 
\cite{Coquereaux2}
\be
{\rm d}(a\otimes A)={\rm d}_{\rm V}a\otimes A+(-1)^{{\rm deg}\,a}\,a
\otimes{\rm d}_{\rm H}A\,,
\ee
or
\be
{\rm d}\pmatrix{A&C\cr D&B}=\pmatrix{{\rm d}_{\rm H}A&-{\rm d}_{\rm H}C\cr
-{\rm d}_{\rm H}D&{\rm d}_{\rm H}B}-m\pmatrix{
{\rm e}^{i\gamma}C+{\rm e}^{-i\gamma}D&i{\rm e}^{-i\gamma}(A-B)\cr
-i{\rm e}^{i\gamma}(A-B)&{\rm e}^{i\gamma}C+{\rm e}^{-i\gamma}D},
\ee
in terms of matrices.

A duality operation on the universal algebra of forms on $M\times{\bf Z}_2$ 
can also be constructed in a straight\-forward manner.  The dual of 
$a\otimes A$ is
\be
\ast(a\otimes A)=(-1)^{\deg a\deg A}\,\ast a\otimes\ast A\,,
\ee
where $\ast$ acting on $A$ denotes the usual Hodge star operation on $M$, 
while that on $a$ is defined by (\ref{star}).  Explicitly, it reads
\be
\ast\pmatrix{A&C\cr D&B}=\pmatrix{\ast A&(-1)^{{\rm deg}\,C}i\ast C\cr
-(-1)^{{\rm deg}\,D}i\ast D&-\ast B}.
\ee
On the other hand, the involution on $M$ and that on ${\bf Z}_2$ extend to 
this case by the formula
\be
(a\otimes A)^\dagger=(-1)^{\deg a\deg A}\,a^\dagger\otimes A^\dagger,
\ee
or, equivalently,
\be
\pmatrix{A&C\cr D&B}^\dagger=\pmatrix{
A^\dagger&(-1)^{{\rm deg}\,D}D^\dagger\cr
(-1)^{{\rm deg}\,C}C^\dagger&B^\dagger}.
\ee
An invariant volume element and inner product on this extended geometry
follow in the usual way.

The calculus developed in this subsection is similar to that of 
ref.~\cite{Coquereaux2}.  There are some minor differences in the above 
formulae, because we start off with a slightly different definition of the 
exterior derivative operator in (\ref{anti-comm}).

\newsubsection{Gauge-invariant action}

We shall construct a generalised Yang--Mills gauge theory on $M$ with 
unitary Lie group ${\cal G}$, tensored with a ${\bf Z}_2$ internal 
structure.  Let us write the combined connection one-form as \cite{Madore1}
\be
\omega={\cal A}+\theta+\phi\,,
\ee
where the horizontal component ${\cal A}$ can be identified as the usual 
(Hermitian) Yang--Mills gauge connection.  The vertical component 
$\theta+\phi$ is the gauge connection corresponding to ${\cal U}$, which 
has already been discussed in sec.~2.3.  Under a generalised gauge 
transformation, we have
\be
\omega~\rightarrow~g^{-1}\omega g-g^{-1}i{\rm d}g\,,
\ee
where $g\in{\cal G}\times{\cal U}$.  This can be decomposed into 
transformations of the individual components of $\omega$:
\ba
\label{gauge_transf1}
{\cal A}~&\rightarrow&~g^{-1}{\cal A}g-g^{-1}i{\rm d}_{\rm H}g\,, \nonumber\\
\theta~&\rightarrow&~g^{-1}\theta g-g^{-1}i{\rm d}_{\rm V}g\,, \nonumber\\
\phi~&\rightarrow&~g^{-1}\phi g\,. 
\ea
Moreover, $\theta$ is invariant under this gauge transformation.  The 
curvature two-form is
\be
\Omega={\rm d}\omega+i\,\omega\wedge\omega
={\cal F}+{\rm D}_{\rm H}\phi+m^2-\phi^2,
\ee
where ${\cal F}={\rm d}_{\rm H}{\cal A}+i\,{\cal A}\wedge{\cal A}$ is the 
usual Yang--Mills curvature, and ${\rm D}_{\rm H}\phi={\rm d}_{\rm H}\phi
+i[{\cal A},\phi]$ the Yang--Mills gauge-covariant derivative of $\phi$.

Now, we shall write
\be
{\cal A}=\pmatrix{A&0\cr0&B},\qquad
\phi=\pmatrix{0&\varphi\cr\varphi^\dagger&0},
\ee
where $A$ and $B$ are one-forms on $M$, and $\varphi$ a scalar field on 
$M$.  These component fields take values in the Lie algebra of ${\cal G}$.  
If we set
\be
g=\pmatrix{g_1&0\cr0&g_2},
\ee
for group elements $g_1,g_2\in{\cal G}$, the gauge transformations 
(\ref{gauge_transf1}) become
\ba
A~&\rightarrow&~g_1^{-1}Ag_1-g_1^{-1}i{\rm d}_{\rm H}g_1\,, \nonumber\\
B~&\rightarrow&~g_2^{-1}Bg_2-g_2^{-1}i{\rm d}_{\rm H}g_2\,, \nonumber\\
\varphi~&\rightarrow&~g_1^{-1}\varphi g_2\,. 
\ea
$A$ is therefore the gauge connection associated with the left action of 
${\cal G}$ on $\varphi$, while $B$ is that associated with the right action 
of ${\cal G}$ on $\varphi$.  The curvature form is \cite{Coquereaux2}
\be
\Omega=\pmatrix{F+m^2-\varphi\varphi^\dagger&-{\rm D}_{\rm H}\varphi\cr
-{\rm D}_{\rm H}\varphi^\dagger&G+m^2-\varphi^\dagger\varphi},
\ee
where $F={\rm d}_{\rm H}A+i\,A\wedge A$, $G={\rm d}_{\rm H}B+i\,B\wedge B$, 
and ${\rm D}_{\rm H}\varphi={\rm d}_{\rm H}\varphi+i(A\varphi-\varphi B)$.  
It can be checked to satisfy the generalised Bianchi identity
\be
\label{Bianchi}
{\rm D}\Omega=0\,,
\ee
where the covariant derivative is, as usual, given by ${\rm D}={\rm d}
+i[\omega,\,\cdot\,]$.

In the notation of tensors, the ${n(n-1)\over2}+n+1$ non-vanishing 
components of $\Omega$ can be taken to be
\ba
&\Omega_{\alpha\beta}=\pmatrix{F_{\alpha\beta}&0\cr0&G_{\alpha\beta}},
&\nonumber\\
&\Omega_{\alpha(n+1)}=\hbox{$\frac{i}{m}$}\pmatrix{{\rm e}^{i\gamma}
{\rm D}_\alpha\varphi&0\cr
0&-{\rm e}^{-i\gamma}{\rm D}_\alpha\varphi^\dagger},&\nonumber\\
&\Omega_{(n+1)(n+2)}=\hbox{$\frac{1}{m^2}$}\pmatrix{m^2
-\varphi\varphi^\dagger&0\cr0&-m^2+\varphi^\dagger\varphi}.&
\ea
In particular, $\Omega_{\alpha(n+2)}=0$.  The norm-square of $\Omega$ is
\be
\hbox{$1\over2$}\int{\rm d}^nx\,{\rm Tr}\,(\Omega_{ij}^\dagger\Omega^{ij})
=\int{\rm d}^nx\left\{\hbox{$\frac{1}{2}$}F_{\alpha\beta}F^{\alpha\beta}
+\hbox{$\frac{1}{2}$}G_{\alpha\beta}G^{\alpha\beta}
+2{\rm D}_\alpha\varphi^\dagger{\rm D}^\alpha\varphi
+2(m^2-\varphi^\dagger\varphi)^2\right\},
\ee
where the integral is over $M$ and trace over the Lie algebra of 
${\cal G}$ is implied. This functional is extremised when the condition
\be
\label{eqn_motion}
{\rm D}\ast\Omega=0\,,
\ee
is satisfied. It translates into the component equations
\ba
{\rm D}_\beta F_{\alpha\beta}&=&i({\rm D}_\alpha\varphi\varphi^\dagger
-\varphi{\rm D}_\alpha\varphi^\dagger)\,,\nonumber\\
{\rm D}_\beta G_{\alpha\beta}&=&i({\rm D}_\alpha\varphi^\dagger\varphi
-\varphi^\dagger{\rm D}_\alpha\varphi)\,,\nonumber\\
{\rm D}_\alpha{\rm D}_\alpha\varphi
&=&-2(m^2-\varphi\varphi^\dagger)\varphi\,.
\ea

Suppose $g_1=g_2$ and $A_\alpha=B_\alpha$.  Following the usual case, we 
may take the action to be $I={1\over4e^2}(\Omega,\Omega)$, where the 
dimensionless parameter $e$ plays the r\^ole of a gauge coupling constant.  
Under the rescaling $\omega\rightarrow e\omega$, it becomes
\be
\label{scalar}
\hbox{$1\over4$}(\Omega,\Omega)=
\int{\rm d}^nx\,\left\{\hbox{$\frac{1}{4}$}F_{\alpha\beta}F^{\alpha\beta}
+\hbox{$\frac{1}{2}$}{\rm D}_\alpha\varphi^\dagger{\rm D}^\alpha\varphi
+V(\varphi)\right\},
\ee
where now $F_{\alpha\beta}=\partial_\alpha A_\beta-\partial_\beta A_\alpha
+ie\,[A_\alpha,A_\beta]$, ${\rm D}_\alpha\varphi=\partial_\alpha\varphi
+ie\,[A_\alpha,\varphi]$, while $\Omega_{(n+1)(n+2)}$ has acquired an 
extra factor of $e$. The potential is
\be
V(\varphi)=-\mu^2\varphi^\dagger\varphi
+\hbox{$\frac{1}{2}$}\lambda(\varphi^\dagger\varphi)^2.
\ee
Here, $\mu=em$ and $\lambda=e^2$ are the two parameters of the theory.  
This familiar action describes Yang--Mills theory coupled to a scalar field 
possessing a quartic self-interaction, with the latter transforming under 
the adjoint representation of ${\cal G}$.  The usual field equations arise 
naturally from (\ref{eqn_motion}).

Observe that the potential acquires a minimum at a non-vanishing value 
$\varphi_0$ of $\varphi$, which satisfies
\be
\varphi^\dagger\varphi=\frac{\mu^2}{\lambda}\,.
\ee
This traces out a sphere of minima in the space of complex $\varphi$, which 
corresponds to an infinity of degenerate vacuum states.  Any particular 
choice of a vacuum state $\varphi_0$ on this sphere however, breaks the 
gauge symmetry.

As anticipated earlier, there is another term which can be added to the 
action, proportional to
\be
|(\theta\wedge\theta,\Omega)|=
em^2\int{\rm d}^nx\,(m^2-\varphi^\dagger\varphi)\,.
\ee
Let us denote this constant of proportionality by $\frac{1}{4}\rho$.  We 
obtain an action again of the form (\ref{scalar}), but with the parameter 
$\mu^2=(e+\frac{1}{4}\rho)em^2$ no longer constrained to being positive.  
When it is negative, the scalar field acquires a mass and its potential has 
a unique minimum.  There is no spontaneous symmetry breaking.

We close this subsection by pointing out when the gauge group ${\cal G}$ is 
Abelian, we have to take $B_\alpha=-A_\alpha$, and so 
$G_{\alpha\beta}=-F_{\alpha\beta}$.  The action remains unchanged, although 
the covariant derivative is now given by ${\rm 
D}_\alpha\varphi=\partial_\alpha\varphi+2ieA_\alpha\varphi$.  These facts 
will be needed when we consider the Abelian Higgs model below.

\newsubsection{Gauge principle underlying Higgs fields}

If the group ${\cal G}$ were trivial in the foregoing analysis, the action
\be
\label{gen_action}
I=\hbox{$\frac{1}{4}$}\left\{(\Omega,\Omega)
+\rho\,|(\theta\wedge\theta,\Omega)|\right\},
\ee
reduces to
\be
\int{\rm d}^nx\,\left\{
\hbox{$\frac{1}{2}$}\partial_\alpha\varphi^\ast\partial^\alpha\varphi
-\mu^2\varphi^\ast\varphi
+\hbox{$\frac{1}{2}$}\lambda(\varphi^\ast\varphi)^2\right\}.
\ee
It describes pure scalar field theory with a quartic self-interaction.  The 
gauge symmetry in question is given by the group ${\cal U}$, which 
translates into the global U(1) symmetry $\varphi\rightarrow{\rm 
e}^{i\Lambda}\varphi$.  When $\rho$ is chosen such that $\mu^2$ is 
positive, we may absorb $\rho$ into $e$ to leave an action purely of the 
form $\frac{1}{4}(\Omega,\Omega)$.  Thus, scalar field theory with a 
symmetry-breaking potential is a gauge theory on $M\times{\bf Z}_2$ with 
the usual quadratic action.  In this subsection, we will show that the 
underlying gauge principle corresponds to the requirement that physics does 
not depend on the choice of vacuum out of a degenerate infinity of possible 
states.

Observe that the curvature form (\ref{curvature-form1}) is minimised when
\be
\phi=-\theta\,.
\ee
$\theta$ thus parametrises the non-trivial vacuum of the theory, through 
the angle parameter $\gamma$ in (\ref{eta}).  As pointed out in sec.\ 2.1, 
it resembles a Maurer--Cartan form.  Recall that in ordinary Yang--Mills 
theory of a Lie group ${\cal G}$, the Maurer--Cartan form can be written as 
$\theta=-h^{-1}i{\rm d}_{\rm H}h$, for an element $h\in{\cal G}$ 
\cite{Eguchi,Madore}.  It satisfies the equation ${\rm d}_{\rm 
H}\theta+i\,\theta\wedge\theta=0$, and so has zero curvature.  In the 
present case however, $\theta$ has a non-vanishing curvature 
(\ref{Maurer}), which is ultimately responsible for the appearance of a 
circular vacuum in the potential, of radius $m$ in the complex-$\varphi$ 
plane.  Fixing a value of $\gamma$ is tantamount to choosing a particular 
vacuum state on this circle, which breaks the U(1) symmetry, since $\theta$ 
is not invariant under the transformation $\theta\rightarrow g^{-1}\theta 
g$.

It therefore follows from (\ref{connection-form}) that $\omega$ describes 
the physical excitation of $\phi$ about a chosen vacuum state $-\theta$.  
Like $\theta$, it does not respect the U(1) symmetry.  However, local 
variations of $\omega$, given by
\be
\delta\omega=-i{\rm D}_{\rm V}\varepsilon=[\phi,\varepsilon]\,,
\ee
for some infinitesimal $\varepsilon\in{\cal U}$, coincide with those of 
$\phi$, as to be expected.

Having identified the physical meanings of the fields $\theta$ and 
$\omega$, let us summarise the general situation.  We first demand a field 
theory of $\phi$ which has the symmetry
\be
\label{old_transf}
\phi~\rightarrow~g^{-1}\phi g\,.
\ee
Suppose $\phi$ has a non-trivial vacuum structure.  We then encounter 
the problem that any choice of a vacuum state $-\theta$ does not respect 
this symmetry.  However, physics has to be independent of the choice of 
vacuum.  To ensure this, we {\it extend\/} the transformation 
(\ref{old_transf}) to one of the form
\be
\label{new_transf}
\omega~\rightarrow~g^{-1}\omega g-g^{-1}i{\rm d}_{\rm V}g\,,
\ee
where ${\rm d}_{\rm V}g=-i[\theta,g]$, and demand that physical quantities 
be invariant under this transformation.  In particular, $\theta$ respects 
the new symmetry.  We may regard a field $\omega$ transforming under 
(\ref{new_transf}) as a fluctuation of $\phi$ about its vacuum state:
\be
\phi=-\theta+\omega\,.
\ee
A field theory of $\phi$ with symmetry (\ref{old_transf}) has thus been 
turned into one of $\omega$ with a symmetry given by (\ref{new_transf}).

Written in this form, the symmetry (\ref{new_transf}) looks uncannily like 
the generic transformation law of a Yang--Mills gauge connection.  Indeed, 
as we have seen, such an interpretation can be realised on some 
non-commutative geometry, where ${\rm d}_{\rm V}$ is the exterior 
derivative operator on this geometry, and $\omega$ a connection one-form.  
{}From $\omega$, we may construct the curvature two-form $\Omega$, which 
transforms as $\Omega\rightarrow g^{-1}\Omega g$.  Gauge-invariant 
candidates for the action of the theory, like $(\Omega,\Omega)$ and 
$|(\theta\wedge\theta,\Omega)|$, then follow immediately.

As we shall only be interested in Yang--Mills--Higgs theory for the rest 
of this paper, our starting action would be given by (\ref{gen_action}), 
with the condition $\rho>-4e$ implicitly assumed.

\newsection{Topological solitons}

\newsubsection{Kinks}

Let us first consider the case $n=1$. Because Yang--Mills theory is 
trivial in one spatial dimension, we are left with a pure scalar field theory
\be
I=\int{\rm d}x\,\left\{
\hbox{$\frac{1}{2}\frac{{\rm d}\varphi^\ast}{{\rm d}x}
\frac{{\rm d}\varphi}{{\rm d}x}$}
+\hbox{$\frac{1}{2}$}e^2(m^2-\varphi^\ast\varphi)^2\right\},
\ee
from an action of the form $\frac{1}{4}(\Omega,\Omega)$.  We may neglect 
the linear term in (\ref{gen_action}), since its presence simply 
corresponds to a rescaling of $e$.  The only two non-vanishing components 
of the curvature form are\footnote{As explained in section 3.1, we set
$\theta^1={\rm d}x$. The forms $\theta^1$ and $\theta^2$ introduced in 
section 2 are, in this case, denoted by $\theta^2$ and $\theta^3$ 
respectively. This shift in the discrete geometry indices generalises
appropriately when we consider cases with higher $n$ below.}
\ba
&\Omega_{12}=-\hbox{$\frac{1}{m}\frac{\rm d}{{\rm d}x}$}({\rm Re}\,
\varphi+i\,{\rm Im}\,\varphi\,\tau_3)\,,& \nonumber\\
&\Omega_{23}=\hbox{$\frac{1}{m^2}$}e(m^2-\varphi^\ast\varphi)\tau_3\,,&
\ea
where we have fixed $\gamma=\frac{\pi}{2}$ for definiteness.

Since this is a three-dimensional system, the usual notion of a self-dual 
or anti-self-dual curvature form does not hold.  But as we shall see, it 
has an appropriate generalisation.  Consider the equation between two-forms:
\be
\label{self-dual1}
\ast\Omega\wedge\theta=\pm\hbox{$\frac{i}{\sqrt[4]{g}}$}\Omega\,,
\ee
where the fourth root of $g$ is present to ensure that both sides of the 
equation have the same physical dimensions.  Solutions to 
(\ref{self-dual1}) would automatically satisfy the field equation 
(\ref{eqn_motion}), provided ${\rm D}\theta=0$.  This condition holds when 
$\varphi$ is real.\footnote{Should we prefer $\varphi$ to be imaginary, the 
appropriate choice of $\gamma$ is 0.  This freedom lies in the choice of 
vacuum.}

Using the relations $\ast(\theta^1\wedge\theta^2)=\theta^3$ and $\ast
(\theta^2\wedge\theta^3)=\frac{1}{\sqrt{g}}\theta^1$, and recalling that 
$\theta=-i\tau_3\theta^2$, we have
\be
\ast\Omega\wedge\theta=-i\Omega_{12}\tau_3\theta^2\wedge\theta^3
-\hbox{$\frac{i}{\sqrt{g}}$}\Omega_{23}\tau_3\theta^1\wedge\theta^2.
\ee
(\ref{self-dual1}) in component form then reads
\be
\Omega_{12}=\mp\hbox{$\frac{1}{\sqrt[4]{g}}$}\Omega_{23}\tau_3\,,
\ee
which is equivalent to the first-order equation
\be
\label{kink-eqn}
\hbox{$\frac{{\rm d}\varphi}{{\rm d}x}$}=\pm e(m^2-\varphi^2)\,.
\ee
If we obtain the equation of motion by varying $I$ with respect to 
$\varphi$, we would find that it is integrable and in fact identical to 
(\ref{kink-eqn}).  It admits the kink (anti-kink) solution \cite{Rajaraman}
\be
\label{kink}
\varphi(x)=\pm m\tanh(em\,x)\,,
\ee
which approaches $\pm m$ as $x\rightarrow\pm\infty$, and so has finite 
energy.

Now, it can be checked that
\be
\sqrt{g}(\ast\Omega\wedge\theta,\ast\Omega\wedge\theta)=
(\Omega,\Omega)\,,
\ee
by explicit expansion in terms of the curvature components.  This identity, 
together with the non-negativity of the inner product 
$(\Omega\mp\sqrt[4]{g}\ast\Omega\wedge\theta, 
\Omega\mp\sqrt[4]{g}\ast\Omega\wedge\theta)$, implies the following lower 
bound for the energy functional:
\be
(\Omega,\Omega)\geq\sqrt[4]{g}|(\Omega,\ast\Omega\wedge\theta)|\,.
\ee
It is saturated for the kink solution (\ref{kink}), and its energy is 
given by
\be
\hbox{$\frac{1}{4}$}\sqrt[4]{g}|(\Omega,\ast\Omega\wedge\theta)|
=\hbox{$\frac{2}{3}$}em^2\left|\varphi(\infty)-\varphi(-\infty)\right|
=\hbox{$\frac{4}{3}$}em^3.
\ee
As this quantity depends only on the asymptotic values of $\varphi$, the 
kink is stable against decay to the vacuum.

We therefore conclude that (\ref{self-dual1}) provides a suitable notion of 
self-duality of $\Omega$.  Like the case of Yang--Mills instantons 
described in the introduction, the energy is topologically bounded from 
below.  This bound is reached when and only when the curvature form is 
self-dual.

\newsubsection{Vortices}

The Yang--Mills--Higgs system in $n=2$ spatial dimensions, known as the 
Abelian Higgs model, describes a scalar field interacting with a U(1) gauge 
field.  Nielsen and Olesen have found finite-energy solutions to this model 
\cite{Nielsen}, which they interpret to be vortices or magnetic flux tubes.  

The action of the Abelian Higgs model is \cite{Jaffe}
\be
\label{Abelian-Higgs}
I=\int{\rm d}^2x\,\left\{\hbox{$\frac{1}{4}$}F_{\alpha\beta}F^{\alpha\beta}
+\hbox{$\frac{1}{2}$}{\rm D}_\alpha\varphi^\ast{\rm D}^\alpha\varphi
+\hbox{$\frac{1}{2}$}\lambda(m^2-\varphi^\ast\varphi)^2\right\},
\ee
for some positive coupling constant $\lambda$. The vortex solutions are 
recovered by solving the field equations, together with the asymptotic 
conditions
\be
\label{bc}
{\rm D}_\alpha\varphi=0\,,\qquad m^2-\varphi^\ast\varphi=0\,,
\ee
for finite energy. 
They have a non-trivial magnetic field $B=F_{12}$, whose total flux
\be
\Phi=\int{\rm d}^2x\,B\,,
\ee
is quantised in units of $\pi/e$, with charge $k$.

In the special case when $\lambda=e^2$, Bogomol'nyi has shown that the 
action functional of the system satisfies \cite{Bogomolnyi}
\be
\label{energy}
I\ge em^2|\Phi|=\pi m^2|k|\,,
\ee
with equality if and only if
\be
\label{bogo}
{\rm D}_1\varphi=\pm i{\rm D}_2\varphi\,,\qquad
B=\pm e(m^2-\varphi^\ast\varphi)\,.
\ee
It has been established that these first-order equations admit solutions 
for each value of integer $k$ \cite{Jaffe}.  They describe $k$ vortices (or 
anti-vortices) in equilibrium, balanced by the nett effect of the repulsive 
magnetic field and the attractive Higgs field.

We would like to recast this model in the language of non-commutative 
geometry.  As we have seen, $I$ is recovered from the action 
(\ref{gen_action}), where the four non-vanishing gauge curvature components 
on this four-dimensional $M\times{\bf Z}_2$ geometry are
\ba
&\Omega_{12}=B\tau_3\,,& \nonumber\\
&\Omega_{13}=-\hbox{$\frac{1}{m}$}{\rm D}_1({\rm Re}\,\varphi
+i\,{\rm Im}\,\varphi\,\tau_3)\,,& \nonumber\\
&\Omega_{23}=-\hbox{$\frac{1}{m}$}{\rm D}_2({\rm Re}\,\varphi
+i\,{\rm Im}\,\varphi\,\tau_3)\,,& \nonumber\\
&\Omega_{34}=\hbox{$\frac{1}{m^2}$}e(m^2-\varphi^\ast\varphi)\tau_3\,.&
\ea
When $\rho=0$, we obtain the critical case $\lambda=e^2$. That the
action is bounded from below is then a consequence of the inequality
\be
\label{inequality}
(\Omega,\Omega)\ge|(\Omega,\ast\Omega)|\,.
\ee
It is minimised when and only when the curvature form is self-dual 
(anti-self-dual):
\be
\label{self-dual2}
\ast\Omega=\pm\Omega\,.
\ee
Such field configurations obey the field equations by virtue of the Bianchi 
identity (\ref{Bianchi}).  In components, the self-duality condition 
becomes
\be
\Omega_{13}=\pm i\Omega_{23}\tau_3\,,\qquad\Omega_{12}=\pm
\hbox{$\frac{1}{\sqrt{g}}$}\Omega_{34}\,,
\ee
which is easily seen to be precisely the equations (\ref{bogo}). Also 
observe that
\ba
(\Omega,\ast\Omega)
&=&\int{\rm d}^2x\,\left\{4eB(m^2-\varphi^\ast\varphi)+2i({\rm D}_1\varphi
^\ast{\rm D}_2\varphi-{\rm D}_2\varphi^\ast{\rm D}_1\varphi)\right\}
\nonumber\\
&=&\int{\rm d}^2x\,\left\{4em^2B+2i\partial_1(\varphi^\ast{\rm D}_2\varphi)
-2i\partial_2(\varphi^\ast{\rm D}_1\varphi)\right\}\,,
\ea
where we have made use of the identity $i({\rm D}_1\varphi^\ast{\rm 
D}_2\varphi-{\rm D}_2\varphi^\ast{\rm D}_1\varphi)=2eB\varphi^\ast\varphi+ 
i\partial_1(\varphi^\ast{\rm D}_2\varphi)-i\partial_2(\varphi^\ast{\rm 
D}_1\varphi)$.  The latter two terms become line integrals at infinity, 
which vanish by the boundary conditions (\ref{bc}).  Thus 
$\frac{1}{4}|(\Omega,\ast\Omega)|=em^2|\Phi|$, and so the inequality 
(\ref{inequality}) is equivalent to that in (\ref{energy}).

The self-duality condition (\ref{self-dual2}) is identical to that for 
Yang--Mills instantons (\ref{duality}), which is not surprising as both are 
four-dimensional systems.  Indeed, this formal similarity between
the two systems was at the heart of how these results were originally
discovered.

\newsubsection{Monopoles}

We move on to $n=3$ spatial dimensions, and again consider a 
Yang--Mills--Higgs action of the form (\ref{gen_action}).  The seven 
non-vanishing components of $\Omega$ can be written as
\ba
&\Omega_{\alpha\beta}=F_{\alpha\beta}\,,&\nonumber\\
&\Omega_{\alpha4}=-\hbox{$\frac{1}{m}$}\pmatrix{{\rm D}_\alpha\varphi&0\cr
0&{\rm D}_\alpha\varphi^\dagger},&\nonumber\\
&\Omega_{45}=\hbox{$\frac{1}{m^2}$}\sqrt{\lambda}
\pmatrix{m^2-\varphi\varphi^\dagger&0\cr0&-m^2+\varphi^\dagger\varphi},&
\ea
for some positive coupling constant $\lambda$. The action then reduces to 
$\frac{1}{4}(\Omega,\Omega)$, and is explicitly
\be
\label{monopole}
I=\int{\rm d}^3x\,\left\{\hbox{$\frac{1}{4}$}F_{\alpha\beta}F^{\alpha\beta}
+\hbox{$\frac{1}{2}$}{\rm D}_\alpha\varphi^\dagger{\rm D}^\alpha\varphi
+\hbox{$\frac{1}{2}$}\lambda(m^2-\varphi^\dagger\varphi)^2\right\}.
\ee

Following the three-dimensional kink case, we take the condition of 
self-duality (anti-self-duality) of $\Omega$ in this five-dimensional 
system to be
\be
\label{self-dual3}
\ast\Omega=\pm\sqrt[4]{g}\Omega\wedge\theta\,,
\ee
both sides of this equation being three-forms.
The field equation (\ref{eqn_motion}) follows if ${\rm D}\theta=0$, 
namely if $\varphi$ is Hermitian.  Now, (\ref{self-dual3}) is equivalent to
\ba
\label{self-dual3a}
\sqrt[4]{g}\Omega_{\alpha\beta}&=&\mp\epsilon_{\alpha
\beta\gamma}\Omega_{\gamma 4}\,, \nonumber\\
\Omega_{45}&=&0\,.
\ea
The former equation implies that
\be
\label{self-dual4}
F_{\alpha\beta}=\pm\epsilon_{\alpha\beta\gamma}
{\rm D}_\gamma\varphi\,,
\ee
while the latter equation ensures the vanishing of the coupling constant 
$\lambda$.

Now, we have the identity
\be
\sqrt{g}(\ast(\Omega\wedge\theta),\ast(\Omega\wedge\theta))
=(\Omega,\Omega)\,.
\ee
As before, that the norm-square of $\Omega\mp\sqrt[4]{g}
\ast(\Omega\wedge\theta)$ is non-negative then implies
\be
(\Omega,\Omega)\geq\sqrt[4]{g}|(\Omega,\ast(\Omega\wedge\theta))|\,,
\ee
with equality if and only if (\ref{self-dual3}) is satisfied. The lower 
bound for the action is explicitly
\be
\label{top_charge}
\hbox{$\frac{1}{4}$}\sqrt[4]{g}|(\Omega,\ast(\Omega\wedge\theta))|
=\int{\rm d}^3x\,\left\{F_{12}{\rm D}_3\varphi-F_{13}{\rm D}_2
\varphi+F_{23}{\rm D}_1\varphi\right\}.
\ee
Using the Bianchi identity $\epsilon_{\alpha\beta\gamma}{\rm D}_\alpha 
F_{\beta\gamma}=0$, and partial integration, the right-hand side can be 
turned into a surface integral on the two-sphere at infinity.  It is thus a 
topological quantity, and is in fact quantised in units of magnetic charge.

It is well-known that the action (\ref{monopole}) admits a monopole 
solution, discovered in approximate form by 't~Hooft \cite{tHooft} and 
Polyakov \cite{Polyakov} for the case of ${\cal G}={\rm SU}(2)$.  In the 
BPS limit $\lambda\rightarrow0$, exact solutions are known 
\cite{Prasad,Bogomolnyi}.  These BPS monopoles satisfy the first-order 
equation (\ref{self-dual4}).  They have energy given by (\ref{top_charge}), 
and are topologically stable.

The latter equation in (\ref{self-dual3a}) actually imposes that one of the 
two discrete dimensions is trivial.  In this sense, BPS monopoles are 
solutions of a generalised four-dimensional gauge theory, while 
't~Hooft--Polyakov monopoles belong to the full glory of the 
five-dimensional system.

\newsubsection{Instantons}

The case of $n=4$ is in many ways similar to the previous one, with eleven
non-trivial curvature components $\Omega_{\alpha\beta}$, $\Omega_{\alpha5}$ 
and $\Omega_{56}$.  The natural generalisation of the self-duality 
(anti-self-duality) condition on $\Omega$ is
\be
\label{self-dual5}
\ast\Omega=\mp\sqrt{g}i\Omega\wedge\theta\wedge\theta\tau_3\,,
\ee
where the extra factor of $\tau_3$ is needed for consistency.  It implies 
the field equation (\ref{eqn_motion}), provided ${\rm D}\theta={\rm 
D}(\theta\tau_3) =0$.  The former ensures that $\varphi$ is Hermitian, 
while the latter that $\varphi$ is anti-Hermitian.  Together, they imply 
that the Higgs field must vanish.  Indeed, (\ref{self-dual5}) in components 
reads
\ba
&F_{\alpha\beta}=\pm\hbox{$\frac{1}{2}$}\epsilon_{\alpha\beta\gamma\delta}
F_{\gamma\delta}\,,&\nonumber\\
&\Omega_{\alpha5}=\Omega_{56}=0\,.&
\ea
Thus, the Higgs sector becomes trivial, leaving a four-dimensional 
Yang--Mills theory whose curvature form $F_{\alpha\beta}$ satisfies the 
familiar self-duality constraint.  Solutions to (\ref{self-dual5}) are but 
Yang--Mills instantons \cite{Belavin}, which were briefly discussed in the 
introduction.

There is another possible definition of self-duality in this case, of the form
\be
\ast\Omega\simeq\pm\Omega\wedge\Omega\,,
\ee
which has previously been considered by various authors 
\cite{Tchrakian,Bais} in the context of six-dimensional Yang--Mills theory.  
However, this over-constrained set of equations does not admit any 
non-trivial finite-energy solutions, as discovered in ref.\ 
\cite{Batenburg} by analysing the asymptotics of these equations.  This 
is a consequence of a virial theorem for Yang--Mills--Higgs solitons that 
we shall prove in the following subsection.

\newsubsection{A virial theorem}

A well-known theorem of Derrick \cite{Derrick} states that there are no 
static soliton solutions to pure scalar field theory, except in one spatial 
dimension where we have the kink solutions, amongst others.  On the other 
hand, Deser \cite{Deser} has shown that there are no such solutions in pure 
Yang--Mills theory, apart from the instanton in four spatial dimensions.  
These non-existence results are due to the attractive nature of scalar 
fields and the repulsive one of Yang--Mills fields respectively.  When the 
two interactions are combined, we would expect these conditions to be 
relaxed.  It turns out that static solitons in Yang--Mills--Higgs theory 
only exist in one, two and three spatial dimensions, a result previously 
proved in refs.\ \cite{Goddard,Jaffe}.  Kinks, vortices and monopoles are 
therefore a complete list of the different types of soliton solutions to 
this theory.  We shall briefly show how this result can be obtained, by 
directly extending Deser's arguments to the generalised Yang--Mills case on 
space-time with a ${\bf Z}_2$ internal structure.

We shall extend the Euclidean Yang--Mills--Higgs system, given by 
(\ref{scalar}), to $(n+1)$-dimensional Minkowski space-time, with signature 
$(-1,1,\dots,1)$ and the space-time index given by $\mu,\nu=0,1,\dots,n$.  We 
also denote by $\alpha,\beta=1,\dots,n$ the spatial index, and 
$i,j=0,\dots,n+2$ the combined space-time and discrete geometry index.  
The generalised energy-momentum tensor is
\be
\Theta_{\mu\nu}=\hbox{$1\over2$}\left\{-\Omega^\dagger_{(\mu}{}^i
\Omega^{\phantom{}}_{\nu)i}+\hbox{$1\over4$}g_{\mu\nu}\Omega^\dagger{}^{ij}
\Omega_{ij}\right\},
\ee
whose trace is equal to the usual energy-momentum tensor of 
Yang--Mills--Higgs theory.

Consider static solutions with finite energy, other than the vacuum.  To 
ensure $\int{\rm d}^nx\,\Theta^{00}\allowbreak<\infty$, we need all 
$\Omega_{ij}$ to fall off faster than $r^{-n/2}$.  This, together with the 
fact that $\partial_0\Theta^{0\alpha}=0$, implies the identity
\be
\label{divergence}
\int{\rm d}^nx\,{\rm Tr}\,\Theta^\alpha{}_\alpha
=\int{\rm d}^nx\,{\rm Tr}\,\partial_\beta(x^\alpha\Theta_\alpha{}^\beta)=0\,.
\ee
Now, it can be checked that
\be
\Theta^\alpha{}_\alpha=\hbox{$1\over2$}\left\{\hbox{$2-n\over2$}
\Omega^\dagger_{\alpha0}\Omega_{\alpha0}+\hbox{$n-4\over4$}
\Omega^\dagger_{\alpha\beta}\Omega_{\alpha\beta}
-\hbox{$n\over2$}\Omega^\dagger_{a0}\Omega_{a0}
+\hbox{$n-2\over2$}\Omega^\dagger_{\alpha a}\Omega_{\alpha a}
+\hbox{$n\over4$}\Omega^\dagger_{ab}\Omega_{ab}\right\}.
\ee
In the static gauge where the solution is time-independent, we have 
$\Omega_{\alpha0}=0$ \cite{Deser}.  A similar argument can be used to show 
that $\Omega_{a0}=0$.  In this gauge, ${\rm D}_0\varphi=ie[A_0,\varphi]$.  
It follows from the equation of motion ${\rm D}_\alpha F_{0\alpha} 
=i({\rm D}_0\varphi\varphi^\dagger-\varphi {\rm D}_0\varphi^\dagger)=i({\rm 
D}_0\varphi^\dagger\varphi-\varphi^\dagger {\rm D}_0\varphi)$ that
\be
{\rm Tr}\left[({\rm D}_0\varphi\varphi^\dagger-\varphi{\rm D}_0
\varphi^\dagger+{\rm D}_0\varphi^\dagger\varphi-\varphi^\dagger{\rm D}_0
\varphi)A_0\right]=0\,,
\ee
or
\be
{\rm Tr}\left[({\rm D_0}\varphi)^\dagger{\rm D_0}\varphi\right]=0\,.
\ee
Thus, ${\rm D_0}\varphi=0$ and we obtain the desired result. 
(\ref{divergence}) reduces to the spatial integral
\be
\label{divergence1}
\int{\rm d}^nx\,{\rm Tr}\left\{\hbox{$n-4\over4$}\Omega^\dagger_{\alpha 
\beta}\Omega_{\alpha\beta}
+\hbox{$n-2\over2$}\Omega^\dagger_{\alpha a}\Omega_{\alpha a}
+\hbox{$n\over4$}\Omega^\dagger_{ab}\Omega_{ab}\right\}=0\,.
\ee

Observe that (\ref{divergence1}) is a weighted sum of positive-definite 
terms whose coefficients depend on $n$.  In the case of pure Yang--Mills 
theory, $\Omega_{\alpha a}$ and $\Omega_{ab}$ both vanish, and so $n=4$ in 
order for (\ref{divergence1}) to hold.  $\Omega_{\alpha\beta}$ vanishes for 
pure scalar field theory, and we require $n<2$ so that the remaining two 
terms have a chance of cancelling each other.  This is another way of 
proving Derrick's theorem, at least for a $\varphi^4$ potential.  In the 
general case, at least one of the coefficients $n-4\over4$ or $n-2\over2$ 
has to be negative.  Non-trivial solutions can therefore exist only when 
$n<4$.

\newsection{Discussion}

By adding a two-dimensional ${\bf Z}_2$ structure to space-time, it is 
possible to recover scalar fields as components of a gauge connection along 
the discrete direction.  In particular, Yang--Mills--Higgs theory in $n$ 
spatial dimensions emerges from a generalised Yang--Mills theory on this 
$(n+2)$-dimensional non-commutative geometry.  It is in this sense that 
kinks are solutions to a three-dimensional gauge theory, vortices to a 
four-dimensional one, and monopoles to a five-dimensional one.

The case of vortices is formally very similar to that of ordinary 
four-dimensional Yang--Mills theory, and the usual self-duality condition 
yields critically coupled vortices.  We have further defined the notion of 
self-dual gauge fields in three and five dimensions, and showed that kinks 
and BPS monopoles obey the respective self-duality equations.

It is possible to generalise the notion of self-duality to six dimensions 
(and even higher), but it imposes the condition that the Higgs sector is 
trivial.  We thus recover ordinary Yang--Mills instantons.  This, we have 
shown, is the consequence of a virial theorem which states that static 
soliton solutions of Yang--Mills--Higgs theory exist only when the number 
of spatial dimensions is $n\leq3$.

This formalism is equally well-suited to pure scalar field theory.  
Consider, for example, the two-dimensional ${\rm CP}^N$ model 
\cite{Rajaraman}.  It consists of $N+1$ complex scalar fields, with a 
${\rm U}(1)$ gauge symmetry, and which are subject to an orthonormality 
condition.  By introducing an auxiliary gauge field, it is possible to 
write the action in the form (\ref{Abelian-Higgs}), where, of course, 
$F_{\alpha\beta}$ vanishes.  Like the Abelian Higgs model, the ${\rm CP}^N$ 
model enjoys many topological properties.  In the case of critical coupling 
$\lambda=e^2$, the energy functional has the same lower bound 
(\ref{inequality}), which is saturated by analytic functions satisfying 
the self-duality condition (\ref{self-dual2}).

The reader may have discerned some connections between the results of this 
paper and certain supersymmetric theories---in particular, the fact that 
self-dual field configurations are precisely those which admit $N=2$
supersymmetric extensions.  Recall that in $N=2$ supersymmetric Yang--Mills 
theory, the chiral multiplet has, in addition to a vector gauge field, 
scalar and pseudoscalar components which can be identified as Higgs fields.  
This provides an alternative superspace unification of Yang--Mills and 
Higgs fields.  The Bogomol'nyi-type bound also appears naturally as a 
consequence of the supersymmetry algebra \cite{Witten}.  What then is the 
relationship between supersymmetry and non-commutative geometry?

In fact, the bosonic part of $N=2$ supersymmetric Yang--Mills theory has 
been shown to arise from considering pure Yang--Mills theory on an ${\cal 
M}\times {\bf Z}_2\times{\bf Z}_2$ geometry \cite{Chen}.  This result 
should not be surprising as two copies of ${\bf Z}_2$ are needed to give 
the scalar and pseudoscalar fields.  The non-commutative geometry we 
consider is embedded in this larger geometry.  Thus our results, at the 
level of the action, should be consistent with $N=2$ supersymmetry.  More 
precisely, it corresponds to the special case when the pseudoscalar field 
is set to zero.  The key difference is, of course, our approach builds upon 
the geometrical foundations of Yang--Mills theory, while that of 
supersymmetry revolves around its algebra.  Our formalism is also a minimal 
one in the sense that just one scalar field and no fermions are required 
for the theory to be self-consistent and exhibit the desired properties.

We hope to have conveyed to the reader, a sense of the power and elegance 
of a conceptual unification of Yang--Mills and Higgs fields, as afforded by 
non-commutative differential geometry.  By using a simple and well-defined 
computational procedure, it is possible to extend some very important 
topological ideas of Yang--Mills theory to the Yang--Mills--Higgs system.  
No doubt, we have only just scratched the surface; the potential of this 
formalism may be further realised in areas such as Chern--Simons models 
\cite{Dunne} and quantum gauge theories \cite{Lee}.

\bigbreak\bigskip\bigskip\centerline{{\bf Acknowledgement}}
\nobreak\noindent
E.T.~wishes to acknowledge helpful discussions with John Madore
and Andrzej Sitarz.

\bigskip\bigskip

{\renewcommand{\Large}{\normalsize}
}
\end{document}